\documentclass[conference]{IEEEtran}
\IEEEoverridecommandlockouts
\usepackage{amsmath,amssymb,amsfonts}
\usepackage{algorithmic}
\usepackage{graphicx}
\usepackage[full]{textcomp}
\usepackage{xcolor}
\usepackage{array, float}
\usepackage{stix}
\usepackage{ragged2e}
\usepackage{titlesec}
\usepackage{booktabs}
\usepackage{fancyhdr}
\usepackage{soul}
\usepackage[english]{babel}
\usepackage[style=ieee, backend=biber, minbibnames=1, maxbibnames=1]{biblatex}
\usepackage[normalem]{ulem}
\usepackage{enumitem}
\usepackage{caption}

\usepackage{tabularx}
\newcolumntype{h}{>{\hsize=2.25\hsize}X}
\newcolumntype{e}{>{\hsize=1.7\hsize}X}
\newcolumntype{g}{>{\hsize=1.5\hsize}X}
\newcolumntype{l}{>{\hsize=1.25\hsize}X}
\newcolumntype{b}{>{\hsize=1.15\hsize}X}
\newcolumntype{x}{X}
\newcolumntype{n}{>{\hsize=.85\hsize}X}
\newcolumntype{m}{>{\hsize=.75\hsize}X}
\newcolumntype{s}{>{\hsize=.5\hsize}X}
\newcolumntype{t}{>{\hsize=.30\hsize}X}

\titlespacing*{\section}
{0pt}{3.0ex plus 1ex minus .2ex}{1.0ex plus .2ex}
\titlespacing*{\subsection}
{0pt}{3.0ex plus 1ex minus .2ex}{1.0ex plus .2ex}

\begin{document}
\captionsetup[figure]{name={Fig.}}

\onecolumn
\null
\vfill
\begin{center}
    \huge{This article has been accepted for publication in the IEEE International Conference on Quantum Computing and Engineering 2025. This is the accepted manuscript made available via arXiv. }
\end{center}
\vfill
\normalsize{© 2025 IEEE. Personal use of this material is permitted. Permission from IEEE must be obtained for all other uses, in any current or future media, including reprinting/republishing this material for advertising or promotional purposes, creating new collective works, for resale or redistribution to servers or lists, or reuse of any copyrighted component of this work in other works.}

\twocolumn
\clearpage

\title{Relative Wavefront Errors in Continuous-Variable Quantum Communication}

\author{Nathan~K.~Long$^{*\dagger}$, 
        ~John~Wallis$^\ddagger$,
        ~Alex~Frost$^\ddagger$,
        ~Benjamin~P.~Dix-Matthews$^\ddagger$,\\
        ~Sascha~W.~Schediwy$^\ddagger$,
        ~Kenneth~J.~Grant$^*$,
        ~Robert~Malaney$^*$ 
\thanks{$^*$NKL, KJG, and RM are with the School of Electrical Engineering and Telecommunications, University of New South Wales, Kensington, NSW, Australia. $^\dagger$NKL is also with the Sensors and Effectors Division of Defence Science and Technology Group, Edinburgh, SA, Australia. $^\ddagger$JW, BPD, AF, and SWS are with the International Centre for Radio Astronomy Research, University of Western Australia, Perth, WA, Australia. The Commonwealth of Australia (represented by the Department of Defence) supports this research through a Defence Science Partnerships agreement.}%
}

\maketitle

\begin{abstract}
     When undertaking continuous-variable quantum key distribution (CV-QKD) across atmospheric channels, strong classical local oscillators (LOs) are often polarization-multiplexed with the weak quantum signals for coherent measurement at the receiver. Although the wavefronts of the quantum signal and LO are often assumed to experience the same distortion across channels, previous theoretical work has shown that they can experience differential distortions, resulting in relative wavefront errors (WFEs). Such errors have previously been shown to limit CV-QKD performance, in some cases leading to zero secure key rates. In this work, for the first time, we provide strong experimental evidence that relative WFEs are present in some circumstances and that standard assumptions in CV-QKD deployments may need to be revisited. In addition, we demonstrate how turbulence can affect the detailed form of the relative WFEs, thereby indicating that long-range links like terrestrial-satellite channels are likely impacted more than short-range terrestrial-only channels. 
     
\end{abstract}

\section{Introduction}

When performing continuous-variable quantum key distribution (CV-QKD) across atmospheric free-space optical channels, a classical local oscillator (LO) is multiplexed with a quantum signal (hereafter called the ``signal''), and then both are transmitted across a channel. At the receiver, the LO is used to measure the electrical field quadratures in the signal. It is often assumed that the distorted spatial phase wavefront of an LO at the receiver is identical to the distorted spatial phase wavefront of the signal (e.g.,~\cite{Villasenor2020}). However, various factors, such as imperfections in optical hardware and photon leakage between signal and LO, can lead to differential signal and LO wavefront distortion (as analyzed in~\cite{Long2025_wfe}), referred to as relative wavefront errors (WFEs). The resulting incoherence between the signal and the LO wavefronts has been shown to have deleterious effects, even leading to null key rates in CV-QKD~\cite{Long2025_wfe}. 

\section{Experimental Results}

\begin{figure}[b]
    \centering
    \includegraphics[trim={0 2.5cm 0 0}, clip, scale=0.75]{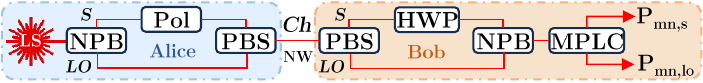}
    \caption{Experimental setup. LS is a laser source, NPB is a non-polarizing beamsplitter, Pol is a polarizer, PBS is a polarizing beamsplitter, Ch is the channel, NW is the nichrome wire, HWP is a half-waveplate, MPLC is a multi-plane light converter, $S$ is the signal path and $LO$ is the LO path. $P_{mn,s}$ and $P_{mn,lo}$ are the signal and LO powers.}
    \label{fig:exp_model}
\end{figure}

Here, we conduct a series of laboratory experiments to expose the presence of relative WFEs in a real setup, whereby a weak signal ($P_{s}=0.16$~$\mu$W) and strong LO ($P_{lo}=0.27$~mW) are generated as fundamental Gaussian modes using the same laser at 1550~nm, then frequency shifted by 1~kHz and polarization-multiplexed, before transmission across a free-space optical channel (see Fig.~\ref{fig:exp_model}). Note, although our weak signal is not a true quantum signal - we make the reasonable assumption that any relative WFE found in our experiment will not be negated when the photon number in the weak signal is reduced. A current is run through a nichrome wire, generating heat using electrical power in the 46~cm long channel for four different conditions, with increasing turbulence strength: Case~0 (no power), Case~1 (power~=~5.40~mW), Case~2 (power~=~12.2~mW), and Case~3 (power~=~21.5~mW). The signal and LO are split at the receiver, aligned in polarization, then measured using a multi-plane light converter (MPLC).

The MPLC decomposes the signal and LO electric fields into the Hermite-Gaussian (HG) basis using a series of phase masks, coupling each mode into a single-mode fiber in the HG$_{00}$ mode (see~\cite{Billault2021}). We measure the first eight modes (at a sample rate of 15~kSa/s) using individual photodetectors, so that the voltage measured, $V_{mn}$, is proportional to the power in each mode, $P_{mn}$ (i.e. ${V_{mn} \propto P_{mn}}$). The signal and LO form a heterodyne beat, so the signal powers~$P_{mn,s}$ and LO powers~$P_{mn,lo}$ can be demultiplexed~\cite{Delange1968}.

To quantify relative WFEs, we take the difference ${\Delta P_{mn}}$ between the normalized signal power ${P_{mn,s}/P_s}$ and the LO power ${P_{mn,lo}/P_{lo}}$ in each mode, ${\Delta P_{mn} = (P_{mn,s}/P_{s}) - (P_{mn,lo}/P_{lo})}$. If the wavefronts of the signal and the LO are the same, then ${P_{mn,s}/P_{s} = P_{mn,lo}/P_{lo}}$, so ${\Delta P_{mn}=0}$, as is commonly assumed. Otherwise, there are relative WFEs.

\begin{figure}
    \begin{centering}
    \includegraphics[scale=0.69]{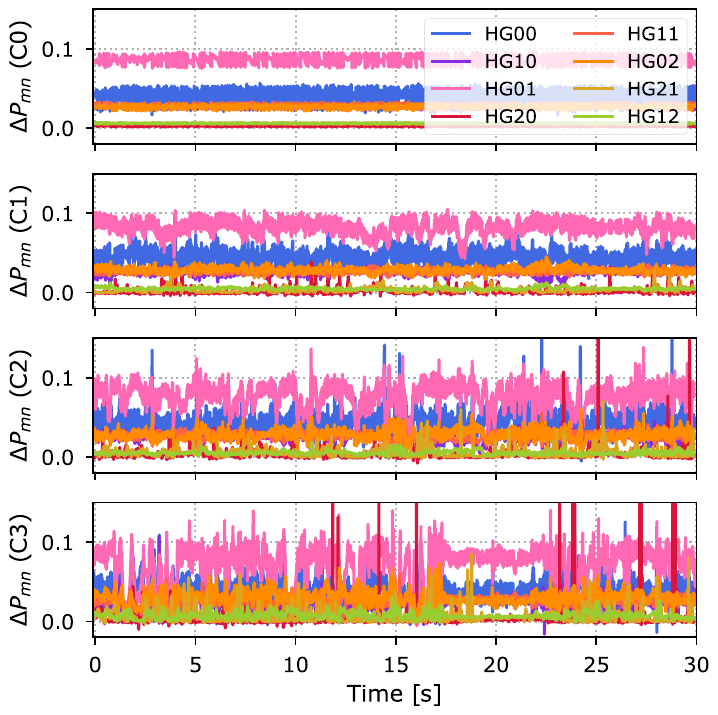}
    \caption{Time versus ${\Delta P_{mn}}$ for turbulence Cases~0-3 (where C represents Case).}\label{fig:dpowers}
    \end{centering}
\end{figure}

Time is plotted against ${\Delta P_{mn}}$ for each of the turbulence cases, over a period of 30~s, in Fig.~\ref{fig:dpowers}. It can be immediately seen that \textit{relative WFEs are present} between the signal and the LO (${\Delta P_{mn}} > 0$) for all cases. Given that relative WFEs are present in Case~0, they are likely caused by imperfections or inaccurate calibration of optical hardware~\cite{Long2025_wfe}. It can be also seen in Fig.~\ref{fig:dpowers} that ${\Delta P_{mn}}$ is highest for the HG$_{01}$ mode, followed by the HG$_{00}$ mode, then the HG$_{10}$, HG$_{11}$, and HG$_{02}$ modes (layered on top of each other). We also see that the HG$_{20}$, HG$_{21}$, and HG$_{12}$ modes approach zero (layered on top of each other). In general, the fluctuations in ${\Delta P_{mn}}$ for each mode remain relatively constant for Case~0, indicating that the WFEs are relatively constant.

\begin{figure}[t]
    \begin{centering}
    \includegraphics[scale=0.7]{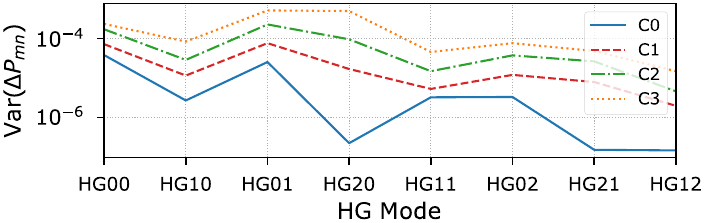}
    \caption{HG mode versus ${\Delta P_{mn}}$ variance for turbulence Cases~0-3 (where C represents Case).}\label{fig:dpowers_var}
    \end{centering}
\end{figure}

From Fig.~\ref{fig:dpowers}, the same ordering of the modes, from the lowest to the highest ${\Delta P_{mn}}$ values, is found for the turbulent Cases~1-3. However, it can be seen that the ${\Delta P_{mn}}$ fluctuations become noisier as the turbulence strength increases, which we quantify in Fig.~\ref{fig:dpowers_var} using the variance of ${\Delta P_{mn}}$. Although the initial WFEs are likely caused by imperfections or imprecise calibration of the optical hardware, the increase in variances of ${\Delta P_{mn}}$ across all modes, as turbulence increases, indicates that the WFEs are affected by the turbulence itself. In addition, we find that the WFEs in each mode are statistically significant by calculating a two-sample Kolmogorov–Smirnov test~\cite{Pratt1981} using the distributions ${P_{mn,s}/P_{s}}$ and ${P_{mn,lo}/P_{lo}}$. Our results show that the differences in all distributions are statistically significant, with a 99.99$\%$ confidence, which supports the existence of relative WFEs.

We note that our experimental CV-QKD setup may not represent all CV-QKD setups, as we adopt path differences for the LO and signal, and apply a small wavelength perturbation for co-measurement of the signal and LO using the MPLC. However, our experiments highlight how relative WFEs may occur in real CV-QKD setups. This result is fundamental to CV-QKD across atmospheric channels; if imperfections in optical hardware and turbulence cause fluctuating relative WFEs, then correcting for them could prove vital in attaining positive key rates across free-space channels. The deleterious effects of relative WFEs on free-space CV-QKD would be particularly pronounced in the future satellite-Earth links required for a global Quantum Internet. To mitigate the effect of WFEs in CV-QKD, our ongoing work proposes a machine learning-based solution to correcting WFEs~\cite{Long2025_wfe}, which can lead to lower excess noise and higher key rates across atmospheric channels.

\section{Conclusion} \label{sec:conc}

We have experimentally demonstrated the presence of relative wavefront errors between a polarization-multiplexed strong classical local oscillator and a weak ``quantum'' signal after passing both through turbulent channels. In addition, we demonstrated how the turbulence strength affects these relative wavefront errors. The presence of relative wavefront errors in CV-QKD setups has the potential to negatively impact QKD secure key rates, necessitating the development of wavefront correction methods to establish a global CV-QKD network.

\printbibliography

\end{document}